# A Neuro-Fuzzy Model with SEER-SEM for Software Effort Estimation

Wei Lin Du

Danny Ho

Luiz Fernando Capretz

University of Western Ontario, London, Canada

{wdu6, dho27, lcapretz}@uwo.ca

## ABSTRACT

Software effort estimation is a critical part of software engineering. Although many techniques and algorithmic models have been developed and implemented by practitioners, accurate software effort prediction is still a challenging endeavor. In order to address this issue, a novel soft computing framework was previously developed. Our study utilizes this novel framework to develop an approach combining the neuro-fuzzy technique with the System Evaluation and Estimation of Resource - Software Estimation Model (SEER-SEM). Moreover, our study assesses the performance of the proposed model by designing and conducting evaluation with published industrial project data. After analyzing the performance of our model in comparison to the SEER-SEM effort estimation model alone, the proposed model demonstrates the ability of improving the estimation accuracy, especially in its ability to reduce the large Mean Relative Error (MRE). Furthermore, the results of this research indicate that the general neuro-fuzzy framework can work with various algorithmic models for improving the performance of software effort estimation.

## 1. INTRODUCTION

The cost and delivery of software projects and the quality of products are affected by the accuracy of software effort estimation. In general, software effort estimation techniques can be subdivided into experience-based, parametric model-based, learning-oriented, dynamics-based, regression-based, and composite techniques [5]. Model-based estimation techniques comprise the general form: $E = a \times Size^b$, where E is the effort, size is the product size, a is the productivity parameters or factors, and b is the parameters for economies or diseconomies [8][19]. In the past decades, some important software estimation algorithmic models have been published by researchers, for instance Constructive Cost Model (COCOMO) [4], Software Life-cycle Management (SLIM) [24], SEER-SEM [9], and Function Points [2][20]. Model-based techniques have several strengths, the most prominent of which are objectivity, repeatability, the presence of supporting sensitivity analysis, and the ability to calibrate to previous experience [3]. On the other hand, these models also have some disadvantages. One disadvantage of algorithmic models is their lack of flexibility in adapting to new circumstances. As a rapidly changing business, the software industry often comes up with new development methodologies, technology and tools, and hence algorithmic models can be quickly outdated. Another drawback of algorithmic models is the strong collinearity among parameters and the complex non-linear relationships between the outputs and contributing factors.

Created specifically for software effort estimation, the SEER-SEM model was influenced by the frameworks of Putnam [24] and Doty Associates [19]. SEER-SEM has two main limitations on effort estimation. First, there are over 50 input parameters related to the various factors of a project. This increases the complexity for managing the uncertainty from the inputs. Second, the specific details of SEER-SEM increase the difficulty of discovering the nonlinear relationship between the parameter inputs and corresponding outputs.

Our study attempts to improve the prediction accuracy of SEER-SEM and resolve the problems caused by the disadvantages of algorithmic models. For accurately estimating software effort, the neural network and fuzzy logic approaches are combined with SEER-SEM. This research is another evaluation for the effectiveness of the general model of neuro-fuzzy with algorithmic model proposed by the previous studies. Published industrial project data was used to evaluate the proposed neuro-fuzzy SEER-SEM model. The data was collected specifically for COCOMO and transferred to the SEER-SEM parameter inputs, utilizing the guides from the University of Southern California (USC) [21]. The estimation performance was finally verified.

## 2. Background
### 2.1 Soft Computing Techniques

Soft computing, which is motivated by the characteristics of human reasoning, has been widely known and utilized since the 1960s. The overall objective from this field is to achieve the tolerance of incompleteness and to make decisions under imprecision, uncertainty, and fuzziness [22][23]. Soft computing has been adopted by many fields, including engineering, manufacturing, science, medicine, and business. The two most prominent techniques of soft computing are neural networks and fuzzy systems. Neural networks have the ability to learn from previous examples, but it is difficult to prove that neural networks are working as expected. Neural networks are like "black boxes" to the extent that the method for obtaining the outputs is not revealed to the users [6][19]. The obvious advantages of fuzzy logic are easy to define and understand an intuitive model by using linguistic mappings and to handle imprecise information [12][18]. On the other hand, it is not easy to guarantee that a fuzzy system with a substantial number of complex rules will have a proper degree of meaningfulness [12]. In addition, the structure of fuzzy if-then rules lacks the adaptability to handle external changes [18]. The obvious strengths of neural networks and fuzzy logic as independent systems have prompted researchers to develop a hybrid neuro-fuzzy system. Specifically, a neuro-fuzzy system is a fuzzy system that is trained by a learning algorithm derived from the neural network theory [22]. Jang's [18] Adaptive Neuro-Fuzzy Inference System (ANFIS) is one type of hybrid neuro-fuzzy system, which is composed of a five-layer feed-forward network architecture.

### 2.2 Soft Computing in Software Effort Estimation

Soft computing is especially important in software cost estimation, particularly when dealing with uncertainty and with complex relationships between inputs and outputs. In the 1990's, a soft computing technique was introduced to build software estimation models and improve prediction performance [7]. Hodgkinson and Garratt [13] introduced the neuro-fuzzy model for cost estimation as one of the important methodologies for developing non-algorithmic models. Their model did not use any of the existing prediction models, as the inputs are size and duration, and the output is the estimated project effort. The clear relationship between Function Point Analysis (FPA) and effort was demonstrated by Abran and Robillard's study [1].

Huang *et al.* [14][15][16] proposed a software effort estimation model that combines a neuro-fuzzy framework with COCOMO II. The parameter values of COCOMO II were calibrated by the neuro-fuzzy technique in order to improve its prediction accuracy. The performance was improved by more than 15% in comparison with that of COCOMO. Xia *et al.* [28] developed a Function Point (FP) calibration model with the neuro-fuzzy technique, which is known as the Neuro-Fuzzy Function Point (NFFP) model. The objectives of this model are to improve the FP complexity weight systems by fuzzy logic, to calibrate the weight values of the unadjusted FP through the neural network, and to produce a calibrated FP count for more accurate measurements. Overall, the evaluation results demonstrated that the average improvement for software effort estimation accuracy is 22%. Wong *et al.* [27] introduced a combination of neural networks and fuzzy logic to improve the accuracy of backfiring size estimates. The study compared the calibrated prediction model against the default conversion ratios. The accuracy of the size estimation only experienced a small degree of improvement.

### 2.3 SEER-SEM Effort Estimation Model

SEER-SEM stemmed from the Jensen software model in the late 1970s, where it was developed at the Hughes Aircraft Company's Space and Communications Group [8][9][19]. In 1988, Galorath Inc. (GAI) started developing SEER-SEM [9], and in 1990, GAI trademarked this model. Over the span of a decade, SEER-SEM has been developed into a powerful and sophisticated model, which contains a variety of tools for performing different estimations that are not limited to software effort. SEER-SEM includes the breakdown structures for various tasks, project life cycles, platforms, languages and applications. Furthermore, the users can select different knowledge bases (KBs) for Platform, Application, Acquisition Method, Development Method, Development Standard, and Class based on the requirements of their projects. There are over 50 parameters that impact the estimation outputs. Among them, 34 parameters are used by the SEER-SEM effort estimation model [10][11]. The SEER-SEM effort estimation is calculated by the following equations:

$$E = 0.393469 \times K \qquad (1)$$

$$C_{tb} = 2000 \times \exp\left(\frac{-3.70945 \times \ln\left(\frac{ctbx}{4.11}\right)}{5 \times TURN}\right) \qquad (2)$$

$$K = D^{0.4} \times \left(\frac{S_e}{C_{te}}\right), C_{te} = C_{tb}/ParmAdjustment \qquad (3)$$

$ctbx =$

$ACAP \times AEXPAPPL \times MODP \times PCAP \times TOOL \times TERM \qquad (4)$

*ParmAdjustment =*
$LANGLEXP \times TSYSTEXP \times DSYSDEXP \times PSYSPEXP \times SIBR$
$REUS \times MULT \times RDED \times RLOC \times DSVL \times PSVL \times RVOL \times SP$
$EC \times TEST \times QUAL \times RHST(HOST) \times DISP \times MEMC \times TIMC$
$\times RTIM \times SECR \times TSVL \qquad (5)$

where,

$E$ is the development effort (in person years),

$K$ is the total life-cycle effort (in person years) including development and maintenance,

$S_e$ is the effective size (SLOC),

$D$ is the staffing complexity,

$C_{te}$ is the effective technology,

$C_{tb}$ is the basic technology.

In SEER-SEM effort estimation, each parameter has sensitivity inputs, with the ratings ranging from Very Low (VLo-) to Extra High (EHi+). Each main rating level is divided into three sub-ratings, such as VLo-, VLo, VLo+. These ratings are translated to the corresponding quantitative value used by the effort estimation calculation.

## 3. A Neuro-Fuzzy SEER-SEM Model

A general soft computing framework for estimation, which is based on the unique architecture of the neuro-fuzzy model described in the patent US-7328202-B2 [17], was built by Huang *et al.* [15]. The framework is composed of inputs, a neuro-fuzzy bank, adjusted values of inputs, an algorithmic model, and outputs for estimation, as depicted in Figure 1. The inputs are rating levels, which can be continuous values or linguistic terms such as Low, Nominal, or High. *V1, …,Vn* are the non-rated values of the algorithmic model. On the other hand, $AI_0, …, AI_m$ are the corresponding adjusted quantitative parameter values of the rating inputs, which are the inputs of the algorithmic model for estimating the final output.

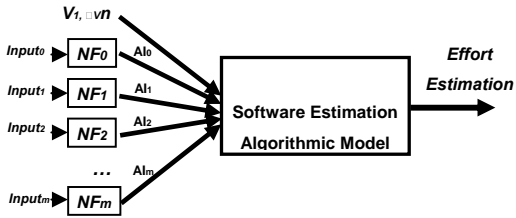

**Figure 1. A General Soft Computing Framework.**

This novel framework has attractive attributes, particularly the fact that it can be generalized to many different situations and can be used to create more specific models. The proposed framework of the neuro-fuzzy model with SEER-SEM, based on the above general structure, is depicted in Figure 2. The inputs include 34 technology and environment parameters, 1 complexity or staffing parameter, and size.

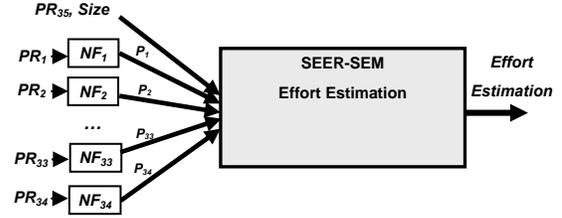

**Figure 2. A Neuro-Fuzzy Model with SEER-SEM.**

$NF_i$ ($i = 1, …, 34$) is a neuro-fuzzy bank, which is composed of thirty-four $NF_i$ sub-models. Through these sub-models, the rating level of a parameter is translated into the corresponding quantitative value ($P_i$, $i = 1, …, 34$) as the inputs of the SEER-SEM effort estimation, as introduced in Section 2.3 from equations (1) to (5). The output of the proposed model is the software effort estimation.

$NF_i$ produces fuzzy sets and rules for training datasets. It translates the rating levels of a parameter into a quantitative value and calibrates the value by using actual project data. Each $NF_i$ uses the structure of the Adaptive Neuro-Fuzzy Inference System (ANFIS), which is a five-layer hybrid neuro-fuzzy system [25], as depicted in Figure 3.

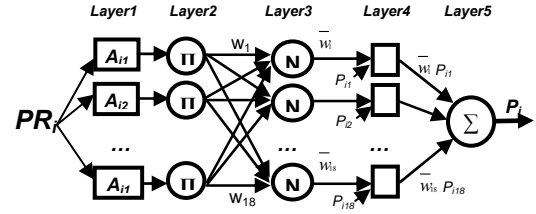

**Figure 3. Structure of $NF_i$.**

- **Functions of Each Layer**

*Layer 1*: In this layer, the membership function of fuzzy set $A$ translates the input, $PR_i$, to the membership grade. The output of this layer is the membership grade of $PR_i$, which is the premise part of fuzzy rules.

$$O_r^1 = \mu_{A_{ir}}(PR_i) \quad \text{for } i = 1, …, 34, r = 1, 2, …, 18 \quad (6)$$

where $O_r^1$ is the membership grade of $A_{ir}$ (=VLo-, VLo, VLo+, Low-, Low, Low+, Nom-, Nom, Nom+, Hi-, Hi, Hi+, VHi-, VHi, VHi+, EHi-, EHi, or EHi+) with the input $PR_i$ or continuous number $x \in [0,19]$; $\mu_{A_{ir}}$ is the membership function of $A_{ir}$.

*Layer 2*: Producing the firing strength is the primary function of this layer. In each node, the label Π multiplies all inputs to produce the outputs according to the defined fuzzy rule for this node. The premise part in the defined fuzzy rule of our proposed model is only based on one condition. Therefore, the output of this layer is the same as the inputs, or membership grade.

$$O_r^2 = w_r = O_r^1 = \mu_{A_{ir}}(PR_i) \qquad (7)$$

**Layer 3:** The function of this layer is to normalize the firing strengths for each node. For each node, the ratio of the $r$th rule's firing strength to the sum of all rules' firing strengths related to $PR_i$ is calculated. The resulting outputs are known as normalized firing strengths.

$$O_r^3 = \overline{w_r} = \frac{w_r}{\sum_{r=1}^{18} w_r} \qquad (8)$$

**Layer 4:** An adaptive result of $P_i$ is calculated with the Layer 3 outputs and the original input of $P_i$ in the fuzzy rules by multiplying $\overline{w_r}$. The outputs are referred to as consequent parameters.

$$O_r^4 = \overline{w_r} P_{ir} \qquad (9)$$

**Layer 5:** This layer aims to compute the overall output with the sum of all reasoning results from Layer 4.

$$O_r^5 = \sum_r O_r^4 = P_i = \sum_r \overline{w_r} P_{ir} \qquad (10)$$

- **Monotonic Constraints**

A monotonic function is a function that preserves the given order. The parameter values of SEER-SEM are either monotonic increasing or monotonic decreasing. Monotonic constraints are used by our model, as a common sense practice, to maintain consistency with the rating levels. For instance, the values of ACAP are monotonic decreasing from VLo- to EHi+, which is reasonable because the higher the analyst's capability, the less spent on project efforts. As for TEST, its values are monotonic increasing because the higher test level causes more effort to be spent on projects.

## 4. Evaluation

### 4.1 Performance Evaluation Metrics

The following evaluation metrics are adapted to assess and evaluate the performance of the effort estimation models.

- **Relative Error (RE)**

$$RE = \frac{(EstimationEffort - ActualEffort)}{ActualEffort}$$

The RE is used to calculate the estimation accuracy.

- **Magnitude of Relative Error (MRE)**

$$MRE = \frac{|EstimationEffort - ActualEffort|}{ActualEffort}$$

- **Mean Magnitude of Relative Error (MMRE)**

$$MMRE = \frac{\left(\sum_{i=1}^{n} MRE_i\right)}{n}$$

The MMRE calculates the mean for the sum of the MRE of n projects. Specifically, it is used to evaluate the prediction performance of an estimation model.

- **Prediction Level (PRED)**

$$PRED(L) = \frac{k}{n}$$

where L is the maximum MRE of a selected range, n is the total number of projects, and k is the number projects in a set of n projects whose MRE <= L. PRED calculates the ratio of a project's MRE that falls into the selected range (L) out of the total projects.

### 4.2 Dataset

For evaluating the neuro-fuzzy SEER-SEM model, over 90 COCOMO data points were collected. Twenty of the 34 SEER-SEM technical parameters can be directly mapped to COCOMO cost drivers and scale factors [26]. The remainder of the SEER-SEM parameters cannot be mapped to COCOMO, and as a result, they are set up as nominal in SEER-SEM.

### 4.3 Evaluation Results

We conducted four studies to evaluate our model. These cases, which used different datasets from the projects, were utilized to perform training on the parameter values. The original SEER-SEM parameter values were trained in each case and the learned parameter values of the four cases were different. We compared the SEER-SEM effort estimation model with our framework. Accordingly, Table 1 presents the MMRE results for Cases 1 to 4. Furthermore, the PRED results are listed under the section for each case. The results for both MMRE and PRED are shown in a percentage format.

Table 1. MMRE Results of all Data Points

| Case ID | SEER-SEM | Validation | Change |
|---------|----------|------------|--------|
| C1 | 84.39 | 61.05 | -23.35 |
| C2 | 84.39 | 59.11 | -25.28 |
| C3 | 84.39 | 59.07 | -25.32 |
| C4-1 | 50.49 | 39.51 | -10.98 |
| C4-2 | 42.05 | 29.01 | -13.04 |

In the tables presenting the analysis results, we have included a column/row named "Change", which is used to indicate the performance difference between the SEER-SEM effort estimation model and our neuro-fuzzy model. For the MMRE, the prediction performance improves as the value becomes closer to zero; therefore, if the change for these performance metrics is a negative value, the MMRE for the neuro-fuzzy model is improved in comparison with SEER-SEM. Additionally, the "PRED(L)" in the later tables represent the prediction level of the selected range. A higher prediction level indicates a greater level of performance for PRED. For PRED, a negative value for the "Change" indicates that our model shows a decreased level of performance as compared to SEER-SEM.

### 4.3.1 Case 1 (C1): Learning with project data points excluding all outliers

This case involved training the parameters of projects where the MREs are lower than or equal to 50%. There are 54 projects that meet this requirement. The learning was done with these 54 project data points, while all data points were used for testing. When using the neuro-fuzzy model, the MMRE decreased from 84.39% to 61.05%, with an overall improvement of 23.35%.

With the neuro-fuzzy model, PRED(20%) and PRED(30%) decreased in comparison to SEER-SEM; however, PRED(50%) and PRED(100%) improved with the neuro-fuzzy model by 7.53% and 10.75% respectively. This indicates that the MRE of the neuro-fuzzy model, in comparison with that of SEER-SEM, contained more outliers that were less than 100% or 50%.

From the results of MMRE and PRED, this calibration demonstrates that the neuro-fuzzy model has the ability to reduce large MREs.

**Table 2. Case 1 PRED Results**

| PRED(L) | 20% | 30% | 50% | 100% |
|---|---|---|---|---|
| SEER-SEM | 36.65% | 45.16% | 56.99% | 81.72% |
| C1 | 29.03% | 37.63% | 64.52% | 92.47% |
| Change | -7.62% | -7.53% | 7.53% | 10.75% |

### 4.3.2 Case 2 (C2): Learning with all project data including all outliers

In Case 2, we used the data points from all projects to calibrate the neuro-fuzzy model without removing the outliers. The testing was performed with the same project dataset used in the training. Using the neuro-fuzzy model, the MMRE decreased by 25.28% in comparison to the MMRE using SEER-SEM. The results of PRED demonstrate that PRED(20%), PRED(30%), and PRED(50%) decreased by more than 20%, while PRED(100%) increased by 16.13% with the neuro-fuzzy model. These results also indicate that the neuro-fuzzy model is effective for improving the MREs that are greater than 100%.

**Table 3. Case 2 PRED Results**

| PRED(L) | 20% | 30% | 50% | 100% |
|---|---|---|---|---|
| SEER-SEM | 36.65% | 45.16% | 56.99% | 81.72% |
| C2 | 15.05% | 18.28% | 36.56% | 97.85% |
| Change | -21.51% | -26.88% | -20.43% | 16.13% |

### 4.3.3 Case 3 (C3): Learning with project data excluding part of outliers

Case 3 calibrated the neuro-fuzzy model by removing the top outliers where the MRE is more than 150%. All data points were used for testing. Overall, as compared to Case 2, calibration excluding the top outliers did not make a significant difference in the performance of the model.

**Table 4. Case 3 PRED Results**

| PRED(L) | 20% | 30% | 50% | 100% |
|---|---|---|---|---|
| SEER-SEM | 36.65% | 45.16% | 56.99% | 81.72% |
| C3 | 15.05% | 18.28% | 38.71% | 97.85% |
| Change | -21.6% | -26.88% | -18.28% | 16.13% |

### 4.3.4 Case 4 (C4): Learning with part of project data points

In Case 4, we used part of the dataset to calibrate the neuro-fuzzy model, and the rest of the data points were used for testing. The objective of this case was to determine the impact of the training dataset size on the calibration results.

- **Case 4 -1 (C4-1): Learning with 75% of project data points and testing with 25% of project data points**

This sub-case performed training with 75% of the project data points and testing with the remaining 25%. In this case, the neuro-fuzzy model improved the MMRE by 10.98%. Furthermore, PRED(30%) and PRED(100%) for our model improved by 4.35% and 8.70% respectively. Finally, with the neuro-fuzzy model, the MREs of all test project data points were within 100%. These results demonstrated the effective performance of the neuro-fuzzy model in reducing large MREs.

**Table 5. Case 4-1 PRED Results**

| PRED(L) | 20% | 30% | 50% | 100% |
|---|---|---|---|---|
| SEER-SEM | 39.13% | 47.83% | 65.22% | 91.30% |
| C4-1 | 34.78% | 52.17% | 60.87% | 100% |
| Change | -4.35% | 4.35% | -4.35% | 8.70% |

- **Case 4 -2 (C4-2): Learning with 50% of project data points and testing with 50% of project data points**

Case 4-2 divided the project data points into two equal subsets, one to train the neuro-fuzzy model and one to perform testing. MMRE improved by 13.04% when using the neuro-fuzzy model. There was no significant difference in the PRED performance from Case 4-1.

**Table 6. Case 4-2 PRED Results**

| PRED(L) | 20% | 30% | 50% | 100% |
|---|---|---|---|---|
| SEER-SEM | 50.00% | 63.04% | 73.91% | 91.30% |
| C4-2 | 43.48% | 56.52% | 76.09% | 100% |
| Change | -6.52% | -6.52% | 2.17% | 8.70% |

### 4.3.5 Evaluation Summary

Figure 4 shows the validation summary for the MMRE across all of the cases. Specifically, the MMRE improves in all of the cases, with the greatest improvement being over 25%.

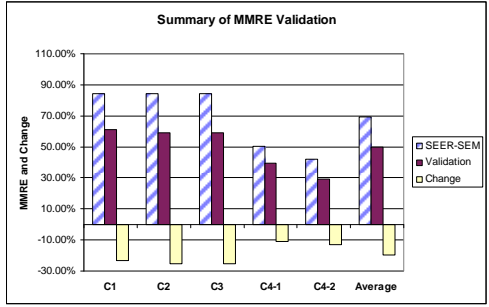

**Figure 4. Summary of MMRE Validation.**

Table 7 illustrates the PRED averages in all of the cases. Compared to the PREDs from SEER-SEM, the averages of PRED(20%), PRED(30%), and PRED(50%) with the neuro-fuzzy model do not show improvement. However, the average of PRED(100%) is increased by 12.14%, which indicates that the neuro-fuzzy model improves the performance of the MMRE by reducing the large MREs.

**Table 7. Summary of PRED Average**

|  | **SEER-SEM** | **Average of Validation** | **Change** |
|---|---|---|---|
| **PRED(20%)** | 39.76% | 27.48% | -12.28% |
| **PRED(30%)** | 49.27% | 36.46% | -12.81% |
| **PRED(50%)** | 62.02% | 55.35% | -6.67% |
| **PRED(100%)** | 85.55% | 97.69% | 12.14% |

## 5. Conclusion and Further Directions

Overall, our research demonstrates that combining the neuro-fuzzy model with the SEER-SEM effort estimation model produces unique characteristics and performance improvements. Effort estimation using this framework is a good reference for the other popular estimation algorithmic models.

The evaluation results indicate that estimation with our proposed neuro-fuzzy model is better than using SEER-SEM alone. In all four cases, the MMREs of our proposed model are improved over SEER-SEM. It is apparent that the neuro-fuzzy technology improves the prediction accuracy. The neuro-fuzzy SEER-SEM model has the advantages of strong adaptability with the capability of learning, less sensitivity for imprecise and uncertain inputs, easy to be understood and implemented, strong knowledge integration, and high transparency. Furthermore, monotonic constraints are used to manage the inputs and outputs as well as the calibrated results.

Although several studies have already attempted to improve the general soft computing framework, there is still room for future work. First, the algorithm of the SEER-SEM effort estimation model is more complex than that of the COCOMO model. The proposed general soft computing framework should be evaluated with even more complex algorithms. Secondly, the datasets in our research are not from the original projects whose estimations are performed by SEER-SEM. When the SEER-SEM estimation datasets are available, more experiments can be run to evaluate the performance of the neuro-fuzzy model.